\documentclass[runningheads]{llncs}
\usepackage[T1]{fontenc}
\usepackage{graphicx}
\usepackage{url}
\begin{document}
\title{Deep Learning-Based Cross-Anatomy CT Synthesis Using Adapted nnResU-Net with Anatomical Feature Prioritized Loss}
\titlerunning{Cross-Anatomy CT Synthesis using Adapted nnResU-Net}
%
\author{Javier Sequeiro Gonzalez\inst{1,2,3}\orcidID{0009-0004-7747-1975} \and
Arthur Longuefosse \inst{1,4}\orcidID{0009-0009-0261-9812} \and
Miguel Diaz Benito\inst{1,2,3}\orcidID{0009-0002-2110-9972} \and Alvaro Garcia Martin\inst{2} \orcidID{0000-0002-1705-3972} \and
Fabien Baldacci \inst{1} \orcidID{0000-0003-4888-4880}} 
\authorrunning{Sequeiro et al.}
%
\institute{Univ. Bordeaux, CNRS, LaBRI, UMR 5800, F-33400 Talence, France \and
Universidad Autónoma de Madrid, Madrid, 28049, Spain
\and 
Pazmany Peter Catholic University, Budapest, 1083, Hungary
\and
RIKEN Center for Integrative Medical Sciences, Medical Data Deep Learning Team, Tokyo, Japan}

\maketitle              

\begin{abstract}
We present a patch-based 3D nnU-Net adaptation for Task 1 (MR-to-CT) and Task 2 (CBCT-to-CT) image translation using the multi-center SynthRAD2025 dataset, covering head and neck (HN), thorax (TH), and abdomen (AB) regions. Our approach leverages two main network configurations: a standard U-Net and a residual U-Net, both adapted from nnU-Net for image synthesis. The Anatomical Feature-Prioritized (AFP) loss was introduced, which compares multi-layer features extracted from a compact segmentation network trained on TotalSegmentator labels, enhancing reconstruction of clinically relevant structures. Input volumes were normalized per-case using z-score normalization for MRIs, and clipping plus dataset-level z-score normalization for CBCT and CT. Training used 3D patches tailored to each anatomical region (e.g. 48×192×224 for TH/AB in Task 1) without additional data augmentation. Models were trained for 1000-1500 epochs, with stochastic gradient descent, a polynomial learning rate schedule, and batch sizes of 2–4, with AFP fine-tuning performed for 500 epochs using a combined L1+AFP objective. During inference, overlapping patches were aggregated via mean averaging with reduced step size of 0.3, and intensities were rescaled to Hounsfield Units by inverting the z-score normalization. Both network configurations were applied across all regions and tasks, allowing consistent model design while capturing local adaptations through residual learning and AFP loss. Qualitative and quantitative evaluation revealed that residual networks combined with AFP yielded sharper reconstructions and improved anatomical fidelity, particularly for bone structures in Task 1 and lesions in Task 2, while L1-only networks achieved slightly better intensity-based metrics. This methodology provides a stable solution for cross-modality medical image synthesis, demonstrating the effectiveness of combining the automatic nnU-Net pipeline with residual learning and anatomically guided feature losses.
\end{abstract}
\section{Introduction}
The SynthRAD2025 Grand Challenge \cite{synthrad2025-challenge} aims to advance the generation of synthetic CT (sCT) images for radiation therapy. The challenge is divided into two key tasks: Task 1 focuses on MR-to-CT synthesis, and Task 2 focuses on CBCT-to-CT synthesis. The main goal is to promote the development of robust and accurate image translation models that can be used clinically. \\
In radiation therapy (RT), accurate treatment planning is critical for effective dose delivery. Currently, computed tomography (CT) is the gold standard because it provides the electron density information necessary for precise dose calculations. However, CT has limited soft tissue contrast, making it difficult to differentiate tumors from healthy tissue. In contrast, magnetic resonance imaging (MRI) offers superior soft tissue contrast, which greatly aids in tumor delineation. While MRI is excellent for target identification, it lacks the quantitative electron density data needed for dose computation. Similarly, cone-beam CT (CBCT) is widely used for patient positioning during treatment but suffers from artifacts that make it unsuitable for direct dose calculation. \\
Image translation from MRI or CBCT to sCT addresses these limitations. By generating an sCT from an MRI, we can leverage the high soft tissue contrast for precise tumor targeting while obtaining the necessary electron density information for dose planning, all without the need for multi-modal registration, which can introduce errors. Similarly, translating a CBCT to a sCT allows for on-the-fly dose verification with reduced artifacts, further improving the RT workflow. This approach ultimately minimizes patient radiation exposure and streamlines the entire treatment planning process. \\
Over the past few years, deep learning (DL) has become the dominant method for medical image translation. Convolutional neural networks (CNNs), with their ability to capture local features, have been a foundational architecture. More recently, vision transformers (ViTs) \cite{vit} and their variants, such as the shifted-windows ViT (Swin-ViT) \cite{swin-vit}, have gained prominence due to their capacity to model long-range dependencies through global self-attention mechanisms. Hybrid models, like the Swin UNETR \cite{swin-unetr}, combine these strengths by using a transformer-based encoder and a CNN-based decoder. One of the most notable frameworks in medical image analysis is nnU-Net, a self-configuring U-Net-like network that has achieved state-of-the-art performance in various segmentation tasks. Its automated design process, which adapts to specific datasets, makes it a highly effective and versatile tool. Recent work has successfully adapted nnU-Net for image translation, showcasing its potential beyond segmentation. \\
In this paper, we present an adaptation of the nnU-Net framework for the image translation tasks of the SynthRAD2025 Grand Challenge. We implemented two models: one using a standard U-Net backbone and another using a Residual U-Net backbone, which we refer to as nnResU-Net. Our approach leverages the self-configuring nature of nnU-Net to automatically optimize key hyperparameters for the specific challenges of MR-to-CT and CBCT-to-CT synthesis. By incorporating a residual architecture, we aim to enhance performance by improving gradient flow and enabling the training of deeper networks. After evaluation, the nnResU-Net model demonstrated superior performance in both tasks, confirming its effectiveness for generating high-quality sCT images and supporting advanced radiation therapy planning.

\section{Data}
We used the SynthRAD2025 Grand Challenge dataset \cite{synthrad-dataset} for both tasks, (MRI-to-CT and CBCT-to-CT image translation. The dataset contained multi-center, multi-manufacturer data from three anatomical regions, namely, head and neck (HN), thorax (TH) and abdomen (AB), acquired under clinical conditions, with scans obtained both with and without contrast enhancement.

\subsubsection*{Registration}
Rigid registration was already performed by the challenge organizers using the \texttt{elastix} toolbox.  
Following the challenge's stage 2 preprocessing pipeline\footnote{\url{https://github.com/SynthRAD2025/preprocessing/}}, deformable registration was applied to further align the input and target images.  
A manual quality control was conducted to remove misaligned image pairs from all datasets, especially in Task~1 where cross-contrast volumes (MR, CT) can lead to mismatches in body contours or anatomical structures. The following numbers of volumes were discarded:  
\begin{itemize}
    \item HN: 21 volumes
    \item AB: 29 volumes
    \item TH: 40 volumes
\end{itemize}
\noindent For Task 2, all volumes were retained for training, as registration usually led to correct alignments in each dataset.

\subsubsection*{Intensity normalization}
MRI inputs were normalized with z-score normalization on a per-case basis. CBCT and CT volumes were clipped to $[-1024, 3071]$ HU, and then z-score normalized using dataset-level statistics (mean, std).

\subsubsection*{Data split}
The training dataset was split into 90\% for training and 10\% for validation. This split was chosen to maximize the number of training cases, rather than using a standard 80/20 approach, as a separate validation dataset was available through the online challenge server.

\section{Model}
\subsection{nnU-Net for translation}
For both tasks (Task 1: MR-to-CT translation, Task 2: CBCT-to-CT translation), we adopted the nnU-Net framework \cite{nnU-net} as the backbone model, adapting it for image-to-image translation rather than segmentation. Originally designed as a self-configuring segmentation pipeline, nnU-Net has consistently achieved state-of-the-art results across a wide range of medical datasets and modalities thanks to its automated preprocessing, architecture adaptation, and training parameter optimization. \\
This work builds upon our previously proposed nnU-Net adaptation \cite{adapted_nnU-net}, which demonstrated that nnU-Net, when configured for synthesis instead of segmentation, can match or surpass more complex GAN-based approaches in cross-modality translation. By removing the discriminator, this adaptation provides stable training dynamics and allows direct integration of custom loss functions tailored to anatomical fidelity. The code is publicly available at \url{https://github.com/Phyrise/nnU-Net_translation}. \\
We investigated three main configurations:
\begin{itemize}
    \item \textbf{Baseline nnU-Net (L1 loss)} --- A standard 3D U-Net automatically configured by nnU-Net, trained with an L1 loss between predicted and reference CT volumes. This model has approximately $\sim30$M parameters.
    \item \textbf{Residual nnU-Net (L1 loss)} --- An adapted version of nnU-Net in which convolutional blocks are replaced by residual blocks. In this residual approach, the network learns only the difference between input and target images rather than the full mapping. This is particularly advantageous for medical image translation, where source and target share high structural similarity, as it allows the network to focus on local modifications instead of relearning shared structures. This model has approximately $\sim57$M parameters.
    \item \textbf{Fine-tuned AFP loss variants} --- Both the baseline and residual nnU-Net models were fine-tuned using a combination of L1 and the Anatomical Feature-Prioritized (AFP) loss \cite{AFP}. The AFP loss computes an L1 distance between multi-layer feature maps extracted from a pre-trained TotalSegmentator model \cite{totalseg} applied to both predicted and ground truth CT volumes. By aligning feature activations from a robust anatomical segmentation network, AFP loss encourages preservation of clinically relevant structures while maintaining global intensity accuracy with the L1 loss.
\end{itemize}
Post-processing consisted solely of patch aggregation. During inference, overlapping patches were reconstructed using mean averaging with a reduced step size of 0.3 (vs.\ default 0.5) to increase overlap and reduce boundary artifacts. Intensities were then rescaled back to Hounsfield Units by inverting the z-score normalization using the initial CT dataset fingerprint.

\subsection{Anatomical Feature-Prioritized (AFP) Loss}

\begin{figure}[ht]
\centering
\begin{tabular}{@{\hspace{0mm}}ccc@{\hspace{0mm}}}
\includegraphics[width=0.32\textwidth]{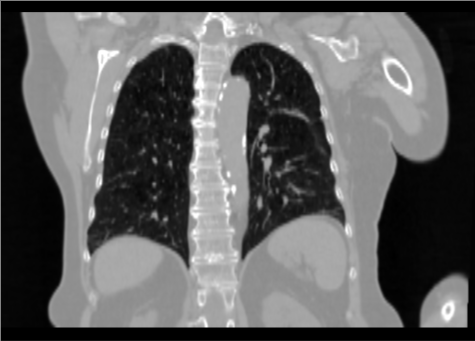} &
\includegraphics[width=0.32\textwidth]{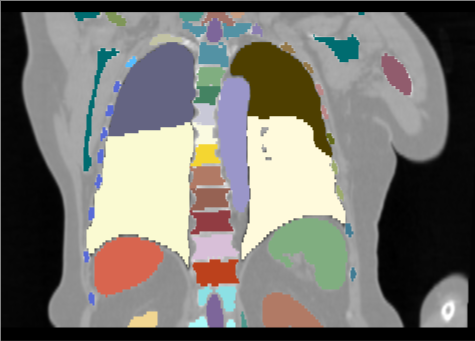} &
\includegraphics[width=0.32\textwidth]{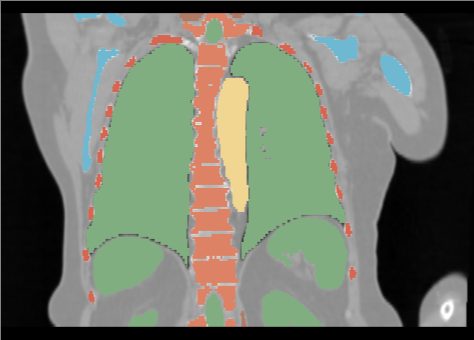} \\
Real CT & TotalSegmentator & TS\_Compact7\_3x1x1
\end{tabular}
\caption{Comparison of axial slices between real CT, segmentation by TotalSegmentator, and segmentation by the adapted TS\_Compact7\_3x1x1 network.}
\label{fig:seg_TS}
\end{figure}

The AFP loss aims to improve the synthesis of anatomically important structures that may be suboptimally reconstructed when relying only on intensity-based losses. Inspired by perceptual loss \cite{perceptual}, AFP computes the L1 distance between feature maps extracted from a 3D segmentation network applied to both predicted and ground truth CT volumes:
\begin{equation}
\mathcal{L}_{\mathrm{AFP}}(x,y) = \frac{1}{N} \sum_{i=1}^{N} \left\| \phi_i(x) - \phi_i(y) \right\|_1
\end{equation}
where $x$ and $y$ denote synthesized and real images, $N$ is the number of extracted layers, and $\phi_i$ represents the feature map from the $i$-th layer of the segmentation network. \\
For the feature extractor, we used TotalSegmentator \cite{totalseg}, widely considered the state-of-the-art for robust segmentation of $>$100 anatomical structures in CT. Two TotalSegmentator configurations are available:  
(1) \textit{standard} --- trained on $1.5\times 1.5\times 1.5$~mm voxels and split into 5 sub-models to handle the large number of labels;  
(2) \textit{fast} --- trained on $3\times 3\times 3$~mm voxels with a single model (used in challenge validation metrics). 
We developed a tailored version for our dataset: CT volumes were segmented using the \textit{standard} TotalSegmentator, and the resulting labels were used to train a new nnU-Net segmentation model adapted to our resolution ($3\times 1\times 1$~mm). Labels were merged into a compact mapping:
\begin{verbatim}
CLASS_MAPPING = {
    "organs":  1,  "cardiac":   2,
    "muscles": 3,  "bones":     4,
    "ribs":    5,  "vertebrae": 6
}
\end{verbatim}
We call this network \texttt{TS\_Compact7\_3x1x1} and use it for AFP feature extraction. This network is designed to be easily trainable by reducing the number of classes from over 100 to a manageable set, is compatible with all three datasets (HN, AB, TH), and is adapted to our pixel spacing so that it can provide meaningful features for the synthesis process using the AFP loss. \\
Figure \ref{fig:seg_TS} illustrates a comparison of axial slices between the original CT from the TH dataset, segmentation by TotalSegmentator, and the corresponding segmentation produced by \texttt{TS\_Compact7\_3x1x1}..

\section{Training}

All models were trained using the nnU-Net training pipeline with modifications for synthesis tasks. The optimizer was stochastic gradient descent (SGD) with an initial learning rate of 0.01, momentum of 0.99, and polynomial learning rate decay. Training was conducted for 1000 epochs for the baseline U-Net model, and 1500 epochs for the Residual U-Net model, with 150 iterations per epoch and a batch size of 4 in both of them. For AFP fine-tuning, the pretrained L1 loss models were further trained for 500 epochs with the combined L1+AFP objective and an updated learning rate of 0.001. The loss weighting was set to $\lambda_{\mathrm{L1}} = 5.0$ to balance intensity fidelity and anatomical preservation. 
Data augmentation was disabled, as preliminary tests showed no substantial improvements; this choice also improves reproducibility and facilitates fairer model comparisons.

\begin{table}[!ht]
\centering
\caption{Patch sizes used for each anatomical region in Task 1 and Task 2.}
\begin{tabular}{c@{\hspace{2em}}c@{\hspace{2em}}c}
\hline
\textbf{Region} & \textbf{Task 1} & \textbf{Task 2} \\ \hline
HN & (56, 192, 192) & (56, 192, 192) \\
AB & (48, 192, 224) & (40, 224, 224) \\
TH & (48, 192, 224) & (40, 224, 224) \\ \hline
\end{tabular}
\label{tab:patch_sizes}
\end{table}
\vspace{-1cm}
\begin{table}[!ht]
\centering
\caption{GPU footprint for different network configurations. nnResU-Net refers to nnU-Net with residual connections. \textit{Ft.} indicates finetuned from a previous model, \textit{ep.} refers to training epoch. All experiments were trained on an NVIDIA A40 GPU.}\label{tab:gpu_footprint}
\begin{tabular}{l @{\hspace{2em}} l @{\hspace{2em}} c @{\hspace{2em}} c @{\hspace{2em}} c @{\hspace{2em}} c}
\hline
\textbf{Model} & \textbf{Loss} & \textbf{Weight Init.} & \textbf{VRAM} & \textbf{Batch} & \textbf{Time per ep.} \\ \hline
nnU-Net    & L1              & Random                & $\sim$20 GB & 4 & 115 s \\
nnU-Net    & L1 + AFP        & Ft. from L1    & $\sim$20 GB & 2 & 80 s \\
nnResU-Net & L1              & Random                & $\sim$20 GB & 4 & 165 s \\
nnResU-Net & L1 + AFP        & Ft. from L1    & $\sim$20 GB & 2 & 100 s \\ \hline
\end{tabular}
\end{table}
\vspace{-0.5cm}
\section{Evaluation}
Model performance was assessed using both quantitative and qualitative metrics. 
Local evaluation employed the same categories of metrics as the challenge:
\begin{itemize}
    \item \textbf{Intensity-based metrics:} Mean Absolute Error (MAE), Peak Signal-to-Noise Ratio (PSNR) and Structural Similarity Index (SSIM).
    \item \textbf{Segmentation-based metrics:} Dice Similarity Coefficient (Dice) and the 95\% Hausdorff Distance (HD95), computed using the \texttt{TotalSegmentator} labels (fast version, trained at $3\times 3\times 3$~mm spacing, identical to the configuration used for the challenge evaluation).
\end{itemize}

\noindent In addition to the quantitative measures, qualitative assessments were performed to visually inspect and compare the reconstruction quality of each model, focusing on anatomical structure preservation and artifact reduction. 

\section{Results}
\subsection{MR to CT translation (Task 1)}
\begin{figure}[!ht]
\begin{tabular}{@{\hspace{0mm}}ccc@{\hspace{0mm}}}
\includegraphics[width=0.32\textwidth]{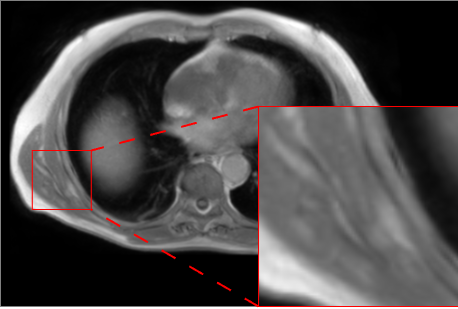}&
\includegraphics[width=0.32\textwidth]{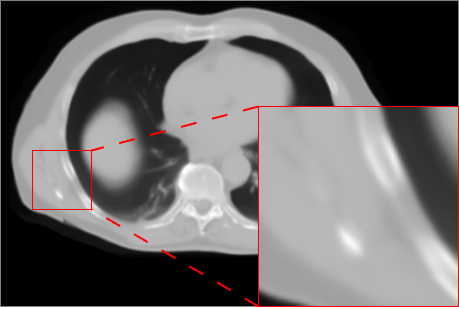}&
\includegraphics[width=0.32\textwidth]{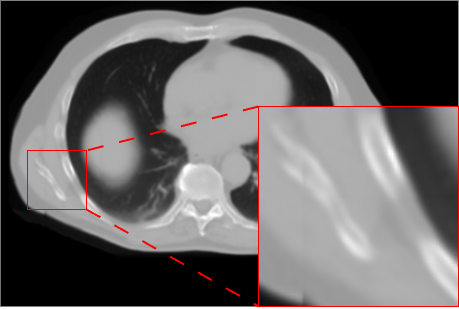} \\
Input MR & nnU-Net L1 & nnU-Net L1 + AFP \\
\includegraphics[width=0.32\textwidth]{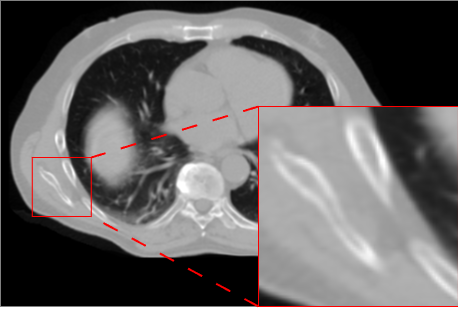}&
\includegraphics[width=0.32\textwidth]{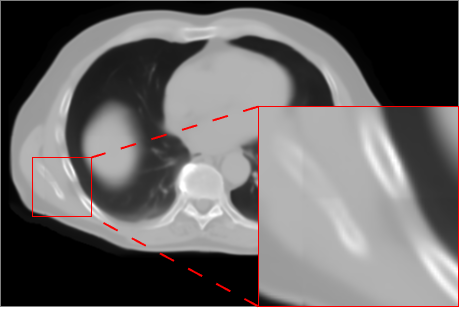}&
\includegraphics[width=0.32\textwidth]{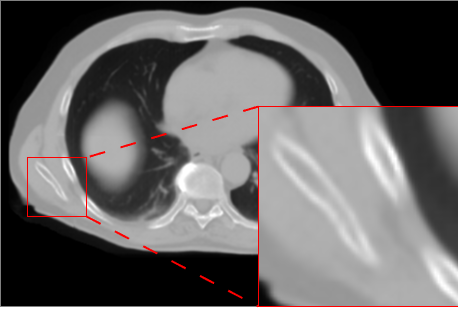} \\
Real CT & nnResU-Net L1 & nnResU-Net L1 + AFP 
\end{tabular}
\caption{Comparison of axial slices between input MR image, ground truth CT, and synthesized CT from several nnU-Net implementations and losses.}
\label{fig:1AB}
\end{figure}

\noindent Figure \ref{fig:1AB} presents qualitative results for the MR-to-CT translation task on the AB dataset. The displayed axial slices include zoomed views on bone structures such as the ribs and right scapula, the latter being among the most challenging to synthesize. Visual inspection reveals that AFP-based models generally produce sharper reconstructions than their L1 counterparts, with nnResU-Net further improving bone delineation compared to nnU-Net. The combination of AFP with nnResU-Net yields particularly accurate results for the scapula, where other models fail to reconstruct the correct morphology. 
\noindent Quantitative results in Table \ref{tab:intensity_results_task_1} and Table \ref{tab:seg_results_task_1} are computed on the combined AB, HN, and TH datasets (three separate models, with all predictions gathered as in the challenge online validation report). For intensity-based metrics, nnResU-Net with L1 loss achieves the best performance across MAE, PSNR, and MS-SSIM, confirming that the residual design improves robustness and gradient flow compared to the baseline nnU-Net. This is consistent with the fact that L1 loss directly optimizes for MAE and related metrics, whereas AFP is not designed to maximize pixel-wise accuracy. In contrast, segmentation-based results show that AFP consistently outperforms L1, and nnResU-Net surpasses nnU-Net, which aligns with the visual observations of more accurate bone reconstruction and sharper anatomical boundaries in AFP-based outputs.

\begin{table}[!ht]
\centering
\caption{Validation server intensity-based results for Task 1.}
\label{tab:intensity_results_task_1}
\begin{tabular}{l @{\hspace{2em}} l @{\hspace{2em}} c @{\hspace{2em}} c @{\hspace{2em}} c}
\hline
\textbf{Model} & \textbf{Loss} & \textbf{MAE} & \textbf{PSNR} & \textbf{MS-SSIM}\\ \hline
nnU-Net    & L1          & $63.908 \pm 23.526$ & $29.730 \pm 2.486$ & $0.9293 \pm 0.0640$ \\
nnU-Net    & L1 + AFP    & $64.982 \pm 23.664$ & $29.676 \pm 2.462$ & $0.9295 \pm 0.0643$ \\
nnResU-Net & L1          & $\mathbf{63.320 \pm 24.126}$ & $\mathbf{29.820 \pm 2.606}$ & $\mathbf{0.9305 \pm 0.065}$ \\
nnResU-Net & L1 + AFP    & $64.349 \pm 24.265$ & $29.743 \pm 2.586$ & $0.9295 \pm 0.066$ \\ \hline
\end{tabular}
\end{table}

\begin{table}[!ht]
\centering
\caption{Validation server segmentation-based results for Task 1.}
\label{tab:seg_results_task_1}
\begin{tabular}{l @{\hspace{2em}} l @{\hspace{2em}} c
@{\hspace{2em}} c}
\hline
\textbf{Model} & \textbf{Loss} & \textbf{DICE} & \textbf{HD95} \\ \hline
nnU-Net    & L1          & $0.7244 \pm 0.1480$ & $9.172 \pm 6.006$ \\
nnU-Net    & L1 + AFP    & $0.7496 \pm 0.1335$ & $7.084 \pm 4.351$ \\
nnResU-Net & L1          & $0.7555 \pm 0.1352$ & $7.629 \pm 4.813$ \\
nnResU-Net & L1 + AFP    & $\mathbf{0.7649 \pm 0.1296}$ & $\mathbf{6.896 \pm 4.285}$ \\
\hline
\end{tabular}
\end{table}

\subsection{CBCT to CT translation (Task 2)}
\begin{figure}[!ht]
\begin{center}
\begin{tabular}{@{\hspace{0mm}}ccc@{\hspace{0mm}}}
\includegraphics[width=0.275\textwidth,  trim=0.2cm 0.3cm 0.3cm 0.3cm, clip]{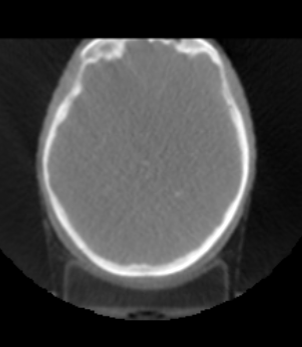}&
\includegraphics[width=0.275\textwidth,  trim=0.2cm 0.3cm 0.3cm 0.3cm, clip]{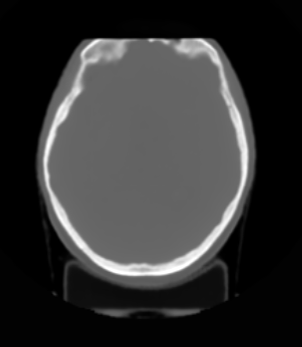}&
\includegraphics[width=0.275\textwidth,  trim=0.2cm 0.3cm 0.3cm 0.3cm, clip]{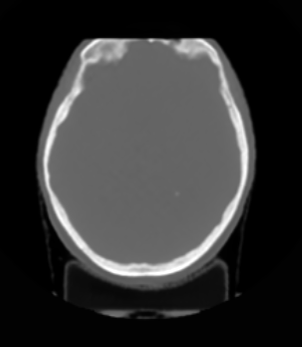} \\
Input CBCT & nnU-Net L1 & nnU-Net L1 + AFP \\
\includegraphics[width=0.275\textwidth,  trim=0.2cm 0.3cm 0.3cm 0.3cm, clip]{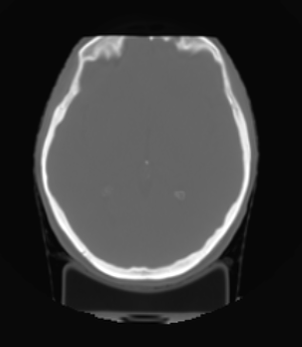}&
\includegraphics[width=0.275\textwidth, trim=0.2cm 0.3cm 0.3cm 0.3cm, clip]{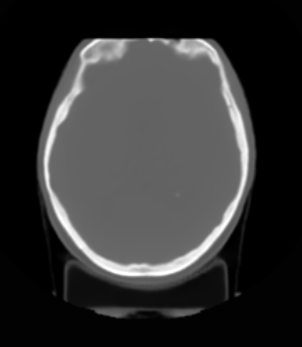}&
\includegraphics[width=0.275\textwidth, trim=0.2cm 0.3cm 0.3cm 0.3cm, clip] {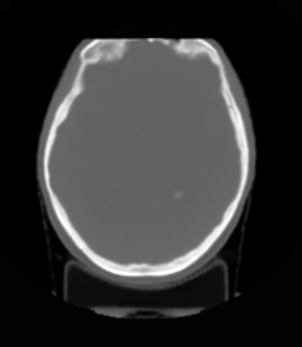} \\
Real CT & nnResU-Net L1 & nnResU-Net L1 + AFP 
\end{tabular}
\caption{Comparison of axial slices between input CBCT image, ground truth CT, and synthesized CT from several nnU-Net implementations and losses.}
\label{fig:2HN}
\end{center}
\end{figure}

\noindent Figure \ref{fig:2HN} presents qualitative results for the CBCT-to-CT translation task on the HN dataset. The displayed axial slices include a lesion visible inside the brain, allowing a focused assessment of reconstruction quality. Visual inspection shows that AFP-based models generally produce slightly sharper and more accurate reconstructions than L1-only models, with nnResU-Net providing the most precise delineation. In particular, the combination of AFP with nnResU-Net better reconstructs the lesion compared to other configurations. 

\noindent Quantitative results in Table~\ref{tab:intensity_results_task_2} and Table~\ref{tab:seg_results_task_2} are computed on the fused HN, AB, and TH datasets. As observed for Task 1, intensity-based metrics are best for nnResU-Net with L1 loss, reflecting the direct optimization of pixel-wise MAE, PSNR, and MS-SSIM. Conversely, segmentation-based metrics show a clear advantage for AFP, with nnResU-Net + AFP achieving the highest DICE and lowest HD95, in line with visual observations of more accurate anatomical boundaries and lesion reconstruction. Overall, the trends across models and losses are consistent with Task 1, confirming the benefits of AFP for improving structure-level fidelity without necessarily maximizing intensity-based metrics.

\begin{table}[!ht]
\centering
\caption{Validation server intensity-based results for Task 2.}
\label{tab:intensity_results_task_2}
\begin{tabular}{l @{\hspace{2em}} l @{\hspace{2em}} c @{\hspace{2em}} c @{\hspace{2em}} c}
\hline
\textbf{Model} & \textbf{Loss} & \textbf{MAE} & \textbf{PSNR} & \textbf{MS-SSIM}\\ \hline
nnU-Net    & L1              & 	$53.342 \pm 14.918$                & $31.679 \pm 2.422$ & $0.962 \pm 0.025$ \\
nnU-Net    & L1 + AFP        & $54.355 \pm 15.006$               & $31.598 \pm 2.414$ & $0.962 \pm 0.025$ \\
nnResU-Net & L1              & $\mathbf{52.958 \pm 14.730}$             & $\mathbf{31.690 \pm 2.454}$ & $\mathbf{0.963 \pm 0.026}$ \\
nnResU-Net & L1 + AFP        & $53.806 \pm 14.911$ & $31.622 \pm 2.454$ & $0.962 \pm 0.026$ \\ \hline
\end{tabular}
\end{table}

\begin{table}[!ht]
\centering
\caption{Validation server segmentation-based results for Task 2.}
\label{tab:seg_results_task_2}
\begin{tabular}{l @{\hspace{2em}} l @{\hspace{2em}} c
@{\hspace{2em}} c}
\hline
\textbf{Model} & \textbf{Loss} & \textbf{DICE} & \textbf{HD95} \\ \hline
nnU-Net    & L1              & $0.814 \pm 0.096$ & $6.0412 \pm 3.939$ \\
nnU-Net    & L1 + AFP        & $0.829 \pm 0.090$ & $5.445 \pm 3.359$ \\
nnResU-Net & L1              &  $0.825 \pm 0.092$ & $5.517 \pm 3.308$ \\
nnResU-Net & L1 + AFP        &  $\mathbf{0.835 \pm 0.088}$ & $\mathbf{5.210 \pm 2.886}$ \\ \hline
\end{tabular}
\end{table}

\section{Discussion}

This study presents a robust baseline for cross-modality medical image translation, specifically MR-to-CT and CBCT-to-CT, based on an adapted nnU-Net framework. Our experiments demonstrate that the approach effectively reconstructs CT volumes across both tasks, with the residual variant (nnResU-Net) consistently outperforming the standard nnU-Net. In terms of loss functions, training nnResU-Net with L1 loss provided superior intensity-based metrics, whereas fine-tuning with the proposed Anatomical Feature-Prioritized (AFP) loss improved segmentation-based performance. From a clinical perspective segmentation accuracy is more relevant than intensity-based metrics. Therefore, our final submission for both tasks were based on nnResU-Net fine-tuned with L1 + AFP losses. During preprocessing, we identified several training pairs with poor alignment, which introduced noise into model training. Applying the edge-based EVOlution registration algorithm \cite{evo-baud} corrected some of these cases; however, since Elastix algorithm is used in the challenge’s deformable registration pipeline for validation and test sets, relying on another registration pipeline risked biasing the results. To avoid this, we opted to retain the original preprocessing pipeline and instead manually remove the most severely misaligned pairs, which ultimately improved model performance. One limitation encountered with AFP loss was the appearance of checkerboard or staircase artifacts, consistent with prior reports on perceptual or feature-based losses \cite{checkerboard}, particularly in standard U-Net decoders. To mitigate this, we replaced transposed convolutions with convolution plus trilinear interpolation, as recommended in \cite{AFP}. Future work will focus on further artifact suppression, potentially by adapting the decoder architecture in nnResU-Net and exploring hybrid configurations such as residual encoders combined with standard U-Net decoders. These strategies aim to preserve the anatomical fidelity benefits of AFP loss while minimizing structural artifacts, thereby improving the clinical reliability of the generated images.

\begin{credits}
\section{Author Contributions}
\noindent
\textbf{Conceptualization}: A.L., F.B., J.S.; 
\textbf{Data curation}: A.L., J.S.; 
\textbf{Formal analysis}: A.L., J.S.; 
\textbf{Investigation}: A.L., J.S., M.D.; 
\textbf{Methodology}: A.L., J.S., M.D; 
\textbf{Software}: A.L., J.S.; 
\textbf{Supervision}: F.B.; 
\textbf{Validation}: A.L., J.S.; 
\textbf{Visualization}: A.L., J.S.; 
\textbf{Writing – original draft}: A.L., J.S.; 

\subsubsection{\ackname} This work was supported by the Video Processing and Understanding Laboratory (VPULab) at Universidad Autónoma de Madrid, which provided the computational resources necessary for the development of this project. The authors would also like to thank Dr. Baudouin Denis De Senneville for his valuable assistance with the registration part of this research.

\subsubsection{\discintname}
The authors have no competing interests to declare that are
relevant to the content of this article.
\end{credits}
%
%
%
%

\end{document}